\documentclass[aps,prl,reprint,twocolumn,oneside]{revtex4-1}
\usepackage{inputenc}
\usepackage{amsmath}
\usepackage{amssymb}
\usepackage{epsfig}
\usepackage{graphicx}

\newcommand{\virg}[1]{``{#1}''}
\newcommand{\gr}[1]{\mathbf{#1}}
\newcommand{\ovl}[1]{\overline{#1}}

\begin{document}

\title{Kinks and waterfalls as signatures of competing order \\ in angle-resolved photoemission spectra of La$_{2-x}$Sr$_x$CuO$_4$}
\author{G. Mazza$^{1,2}$, M. Grilli$^{2,3}$, C. Di Castro$^{2,3}$, and 
S. Caprara$^{2,3}$}
\affiliation{$^1$ International School for Advanced Studies (SISSA), Via Bonomea 
265 34136 Trieste, Italy\\
$^2$CNISM and Dipartimento di Fisica, Universit\`a  di Roma \virg{La 
Sapienza} Piazzale Aldo Moro 5, I-00185 Roma, Italy\\
$^3$ Consiglio Nazionale delle Ricerche, Istituto dei Sistemi Complessi, 
via dei Taurini, I-00185 Roma, Italy}

\begin{abstract} 
We show that the so-called {\it kinks} and {\it waterfalls} observed in angle-resolved 
photoemission spectra of La$_{2-x}$Sr$_x$CuO$_4$, a prototypical high-$T_c$ 
superconducting cuprate, result from the coupling of quasiparticles with {\it two} 
distinct nearly critical collective modes with finite characteristic wave vectors, 
typical of charge and spin fluctuations near a stripe instability. Both phonon-like 
charge and spin collective modes are needed to account for the kinked quasiparticle 
dispersions. This clarifies the long-standing question whether kinks are due to phonons 
or spin waves and the nature of the bosonic mediators of the electron-electron 
effective interaction in La$_{2-x}$Sr$_x$CuO$_4$. 
\end{abstract}
\maketitle


The metallic phase of high-$T_c$ superconducting cuprates evolves remarkably with 
changing the temperature $T$ and the doping $x$. A normal Fermi-liquid behavior is only 
found in overdoped samples, with $x$ larger than the optimal value $x_{\mathrm{opt}}$ 
(where the maximum superconducting critical temperature $T_c$ is achieved). At 
$x\approx x_{\mathrm{opt}}$ and $T>T_c$ the metallic phase seems to be ruled by the 
temperature as the only relevant energy scale, a typical signature of quantum 
criticality. In underdoped samples, with $x<x_{\mathrm{opt}}$, an even stronger anomaly 
is found, with a pseudogap opening around the Fermi energy, below a doping dependent 
temperature $T^*(x)$. Whether this is accompanied by the onset of some sort of ordering 
is still matter of debate. Nonetheless, models with nearly critical collective modes 
(CMs) coupled to fermion quasiparticles (QPs) may not only explain the anomalous 
metallic phase, but also provide candidate mediators of a retarded pairing interaction 
(the so-called {\em glue}) \cite{anderson,scalapino,hanke}, alternative to phonons in 
ordinary superconductors, and are therefore actively investigated. Various proposals
for sources of nearly critical CMs include the antiferromagnetic phase at $x\approx 0$ 
\cite{abanov}, time-reversal-breaking plaquette currents \cite{varma}, order parameters 
with exotic wave symmetry \cite{benfatto,metzner}, or stripe ordering 
\cite{kivelson,CDG}.

In this Letter, we show that the so-called {\it kinks} and {\it waterfalls} 
\cite{garcia} observed in angle-resolved photoemission spectroscopy (ARPES) identify 
charge (C) and spin (S) CMs on the verge of a stripe instability as the main source of 
scattering in La$_{2-x}$Sr$_x$CuO$_4$ (LSCO), solving the long-standing phonon-vs-spin 
issue, at least in LSCO. The sudden changes in the QP velocity (kinks) occurring at 
different energies in different regions of the Brillouin Zone (BZ), have often been 
attributed to a phonon with a near frequency \cite{bogdanov,zhou}, and it has been argued that S 
fluctuations could account for the kinks in YBa$_2$Cu$_3$O$_{7-x}$ (YBCO) 
\cite{scalapino1}, and 
Bi$_2$Sr$_2$CaCu$_2$O$_8$ (BSCCO) \cite{norman1}. We show that the same CMs also account for the 
sudden, nearly vertical, drops of the QP dispersions at high/moderate binding energies 
(waterfalls) \cite{graf,chang}.

A previous survey of Raman spectra \cite{noiraman} for $x=0.15-0.26$ and
various $T$ provided evidence for {\em two} distinct CMs peaked at finite 
characteristic wave vectors in LSCO \cite{notavandermarel}. One CM, 
essentially propagating and centered at typical phonon frequencies, is associated with 
C fluctuations (strongly mixed with the lattice degrees of freedom) near an 
incommensurate charge-density wave instability. The other CM, more diffusive and 
extending to higher energies, is associated with S fluctuations peaked near the wave 
vector of antiferromagnetic order. The behavior of the characteristic low-energy scale 
of the two CMs suggests that a quantum critical point occurs at $x_{QCP}\approx 0.19$ 
\cite{CDGZP,CDGJCPS,loram}, associated to a phase with stripe-like C and S modulation, 
whose onset occurs via a harmonic incommensurate charge-density wave at 
$T\approx T^*(x)$ \cite{CDG,perali,andergassen}.  Remarkably, the strengths 
of the two CMs have an opposite $x$ dependence (see Fig. 5 in Ref.
\cite{noiraman}): The strength of the S CM decreases 
with increasing $x$ and almost vanishes in the most overdoped sample ($x=0.26$), 
whereas the strength of the C CM increases with increasing $x$ and tends to saturate 
in the overdoped regime. The value $x\approx 0.19$ marks 
the boundary between the spin- and the charge-dominated region. At small $x$, S 
fluctuations are naturally enhanced by incipient antiferromagnetism, whereas in the 
optimally and overdoped regime C fluctuations dominate. 

Here, we phenomenologically proceed to analyze the implications of the same two CMs, as 
derived from Raman experiments, on QP spectra. We consider the general Gaussian form of 
CM propagator 
\begin{equation}
\mathcal{D}_{\lambda}(\gr{q},\omega_n)= -\frac{1}{\Upsilon_{\lambda}(\gr{q}
-\gr{Q}_{\lambda}) +|\omega_n| + \omega^2_n/ \ovl{\Omega}_{\lambda}},
\label{eq:CM_periodic}
\end{equation}
where $\lambda=C,S$, $\omega_n$ is the bosonic Matsubara frequency, 
$\Upsilon_{\lambda}(\gr{q})=m_{\lambda}+\nu_{\lambda}\left[2-\cos(q_x a)-\cos(q_y a) 
\right]$ describes the dispersion of a lattice periodic CM, and reproduces the behavior 
$\Upsilon_\lambda(\gr{q}-\gr{Q}_{\lambda}\approx 0)\approx m_\lambda+\frac{1}{2}
\nu_\lambda(\gr{q}-\gr{Q}_{\lambda})^2$ obtained in different contextes for C 
\cite{CDG} and S \cite{abanov,mmp} CMs. $m_{\lambda}$ is proportional to the 
inverse squared correlation length $\xi_\lambda^{-2}$, $\nu_{\lambda}$ sets the 
curvature at the bottom of the CM dispersion law, and $a$ is the spacing of the 
two-dimensional square lattice describing the CuO$_2$ planes of cuprates, henceforth 
taken as unit length. The propagator (\ref{eq:CM_periodic}) is peaked at a 
characteristic wave vector $\gr{Q}_{\lambda}$, has a diffusive character at low energy, 
and becomes more propagating above the energy scale $\ovl{\Omega}_{\lambda}$. The 
CM dispersion is limited by an energy cutoff $\Lambda_{\lambda}$, setting a momentum 
cutoff $|\ovl{\gr{q}}_{\lambda}|\approx(\Lambda_{\lambda}/\nu_{\lambda})^{1/2}$. 
The values of the characteristic wave vectors, $\gr{Q}_C$ and $\gr{Q}_S$, are extracted 
from neutron scattering experiments: The incommensurability of the S density modulation, 
half of the incommensurability of the C density modulation when observed at $x=1/8$ 
\cite{tranquada,abbamonte}, saturates for $x>1/8$, yielding \cite{yamada}
$\gr{Q}_C=\pi\left(\pm\frac{1}{2},0\right),\pi\left(0,\pm\frac{1}{2}\right)$,
$\gr{Q}_S=\pi\left(1\pm\frac{1}{4},1\right),\pi\left(1,1\pm\frac{1}{4}\right)$.

We adopt for the fermion QPs on the CuO$_2$ planes of LSCO a tight-binding dispersion law
including nearest ($t=400$ meV) and next-to-nearest ($t'=-0.21t$) neighbor hopping terms,
\begin{equation}
\epsilon_{\gr{k}}=-2t\left(\cos k_x+\cos k_y\right) -4t'\cos k_x \cos k_y-\mu,
\label{ek}
\end{equation}
where $\mu$ is the chemical potential. Similarly to the electron-phonon coupling, 
QPs are here coupled to CMs through dimensional coupling constants $g_{\lambda}$. 
The survey of Raman data on LSCO \cite{noiraman} yielded the doping evolutions of $m$, 
$\Lambda$, $\ovl\Omega$, and $\kappa\equiv g^2/t\nu$, reported in Tab. \ref{tab}. 
Our aim is to fit ARPES data with the {\em same} CM parameters, although it should be 
borne in mind that Raman response is a momentum integrated quantity, so that the precise 
$\gr{q}$ dependence of the CM-mediated effective interaction, and therefore the value 
of $\ovl{\gr{q}}$, is not fully constrained. We adjust $\ovl{\gr q}$ as a fitting 
parameter \cite{notaqbarra}, which in turn fixes $\nu=\Lambda/\gr{\ovl{q}}^2$ and $g=\sqrt{\kappa t \nu}$. 
The obtained values (see Tab. I), yield four peaks in the CM dispersion, consistent 
with the observation of four separate peaks for the S CM in neutron scattering 
experiments \cite{tranquada,yamada}. 

\begin{table}
\begin{tabular}{l*{7}{c}}
\small{Charge}& & & & &\\
\hline\noalign{\smallskip}
\,~$x$ & $\displaystyle{{m}\atop{\mathrm{(meV)}}}$ & 
$\displaystyle{{\Lambda}\atop{\mathrm{(meV)}}}$ & 
$\displaystyle{{\ovl{\Omega}}\atop{\mathrm{(meV)}}}$ & 
$\kappa$ & $\gr{\ovl{q}}$ & $\displaystyle{{\nu}\atop{\mathrm{(meV)}}}$ & 
$\mathcal{Q}$     \\
\hline\hline\noalign{\smallskip}
0.15 & 2.5  & 248.0 & 25.0 & 5.5 & 0.9 & 300.0 & 0.6   \\
0.17 & 4.35 & 248.0 & 25.0 & 8.0 & 0.9 & 300.0 & 0.6   \\
0.20 & 8.7  & 310.0 & 25.0 & 11.7& 1.0 & 300.0 & 0.5   \\
0.25 & 9.9  & 186.0 & 41.3 & 13.7& 0.9 & 240.0 & 0.5   \\
0.26 & 7.55 & 300.0 & 41.3 & 17.5& 1.0 & 280.0 & 0.55  \\
\hline\noalign\\
\small{Spin}& & & & &\\
\hline\noalign{\smallskip}
\,~$x$ & $\displaystyle{{m}\atop{\mathrm{(meV)}}}$ & 
$\displaystyle{{\Lambda}\atop{\mathrm{(meV)}}}$ & 
$\displaystyle{{\ovl{\Omega}}\atop{\mathrm{(meV)}}}$ & 
$\kappa$ & $\gr{\ovl{q}}$ & $\displaystyle{{\nu}\atop{\mathrm{(meV)}}}$ & 
$\mathcal{Q} $    \\
\hline
\hline\noalign{\smallskip}
0.15 & 0.62 & 86.8 & 248.0 & 4.35 & 0.57 & 260.0 & 0.75  \\
0.17 & 0.62 & 74.4 & 310.0 & 4.5  & 0.6  & 200.0 & 0.6   \\
0.20 & 0.74 & 86.8 & 496.0 & 1.4  & 0.57 & 270.0 & 0.95  \\
0.25 & 1.25 & 62.0 & 496.0 & 0.26 & 0.5  & 240.0 & 1.0   \\
0.26 & 1.5  & 49.6 & 155.0 & 0.40 & 0.42 & 280.0 & 0.85  \\
\hline
\end{tabular}
\caption{Parameters of the C and S CMs:
$m$, $\Lambda$, $\ovl{\Omega}$, and $\kappa$ are extracted from Raman 
data \cite{noiraman}.}
\label{tab}
\end{table} 
 
The effect of CMs on QP spectra is captured computing the lowest-order QP self-energy 
$\Sigma(\gr{k},\omega)=\Sigma_C(\gr{k},\omega)+\Sigma_S(\gr{k},\omega)$, and the 
single-particle spectral density
\[
A(\gr{k},\omega)=\frac{1}{\pi}\frac{\left \vert \text{Im}\Sigma (\gr{k},\omega) \right 
\vert}
{\left[\omega-\epsilon_{\gr{k}}-\text{Re}\Sigma(\gr{k},\omega)\right]^2+
[\text{Im}\Sigma(\gr{k},\omega)]^2}.
\]
The imaginary part of the self-energy is 
\begin{eqnarray}
\nonumber
&&\text{Im}\Sigma_{\lambda}(\gr{k},\omega)=g_{\lambda}^2
\int_{BZ}\frac{d^2 \gr{q}}{(2 \pi)^2}\,
\gamma_{\mathcal Q_\lambda}(\gr{q}-\gr{Q}_\lambda)\\ 
&\times&
\frac{\left(\omega-\epsilon_{\gr{k-q}}\right)\left[f_+(\epsilon_{\gr{k-q}})+
f_-(\epsilon_{\gr{k-q}}-\omega)\right]}
{\left[\Upsilon_{\lambda}(\gr{q}-\gr{Q}_{\lambda})-(\omega-\epsilon_{\gr{k-q}})^2/
\ovl{\Omega}_{\lambda}\right]^2+\left(\omega-\epsilon_{\gr{k-q}}\right)^2},~~~~~
\label{eq:Im_Sigma}
\end{eqnarray}
where a sum over the star of equivalent wave vectors $\gr{Q}_\lambda$ is
understood, $f_\pm(\omega)=1/\left(e^{\omega/T}\pm 1\right)$, and the smooth 
cutoff function $\gamma_{\mathcal Q}(\gr{q})=\exp\left[-\left(2-\cos q_x-
\cos q_y\right)/{\mathcal Q}^2\right]$ accounts for the suppression of the CM-QP 
coupling away from $\gr{Q}_\lambda$. As an order-of-magnitude estimate, 
$\mathcal Q_\lambda\approx |\ovl{\gr{q}}_\lambda|$. For any given $\gr{k}$ and 
$\omega$, we numerically integrate Eq. (\ref{eq:Im_Sigma}) and obtain 
$\text{Re}\Sigma$ via Kramers-Kronig transformation. The chemical potential $\mu$ 
is fixed imposing that 
\[
2\int_{BZ} \frac{d^2\gr{k}}{(2 \pi)^2} \int_{-\infty}^{+\infty} 
d \omega\, A(\gr{k},\omega) f_+(\omega)=1-x.
\]
We calculate the ARPES intensity convoluting $A(\gr{k},\omega)$ with a Gaussian of 
width $\approx 10$\,meV, mimicking energy resolution, and considering only the occupied 
states. 

\begin{figure}[h]
\includegraphics[scale=0.3]{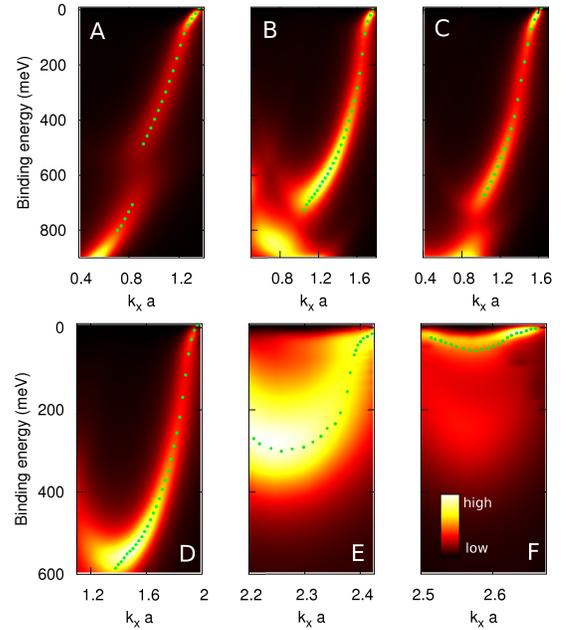}
\caption{(Color online) Spectral density along the cuts of the BZ reported in Fig. 
\ref{fig:self-hs}, for $x=0.15$ and $T=40$\,K. Dots represent the maxima of the MDCs.}
\label{fig4-kink-wf}
\end{figure}

\begin{figure}
\includegraphics[scale=0.3,angle=0]{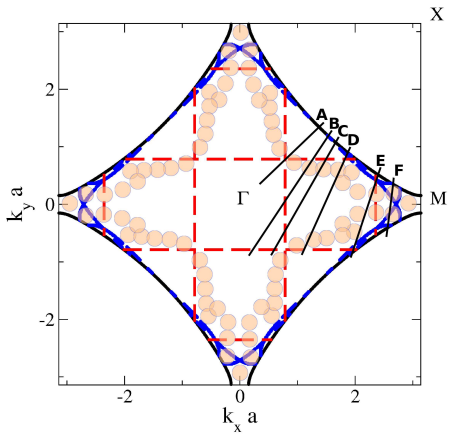}
\caption{(Color online) Fermi surface of LSCO at $x=0.15$ (black solid line). The dashed 
(red online) and solid (blue online) lines mark the C and S hot lines, respectively. 
The shaded circles mark the loci of the waterfalls (from Ref. \cite{chang}). 
The spectra in Figs. \ref{fig4-kink-wf},\ref{fig:spectra} are calculated along the cuts A-F.}
\label{fig:self-hs}
\end{figure}

The main features of ARPES data are determined by the dynamical structure of the 
CMs of Eq. (\ref{eq:CM_periodic}). In Fig. \ref{fig4-kink-wf}, we report the spectra 
along the cuts A-F of Fig. \ref{fig:self-hs}, for $x=0.15$ and $T=40$\,K. We track 
the QP dispersions (dots) as the maxima of the momentum distribution curves (MDCs). 
The condition of quasi criticality (small $m_\lambda$) and the finite values
of the characteristic wave vectors $\gr{Q}_\lambda$ render the scattering most
relevant when the QP momenta $\gr{k}$ satisfy the \emph{hot line} condition 
$\epsilon_{\gr{k}}=\epsilon_{\gr{k} \pm \gr{Q}_\lambda}$, identifying  
the points at the same energy on the QP bands, connected by the characteristic 
wave vectors. 
We report  in Fig. \ref{fig:self-hs} the hot lines for C [(red) dashed lines]
and  S [(blue) solid lines] CMs. These lines intersect the 
Fermi surface at the so-called {\it hot spots}. Away from the hot lines, 
the scattering in Eq. \ref{eq:Im_Sigma} is not dominated by the nearly critical
(low-energy) CM spectrum, and is rather mediated by the whole dynamical range of 
the order of
$\sqrt{\ovl{\Omega}\Lambda}$ \cite{notaqbarra}. This determines the energy scale of 
the kinks appearing at low binding energy ($\lesssim 70$\,meV) in 
Fig. \ref{fig4-kink-wf}. The strong scattering near the hot lines, reminiscent 
of the Bragg scattering occurring when some ordering takes place at specific 
$\gr{Q}_\lambda$, gives instead rise to the high-energy waterfall features 
in Fig. \ref{fig4-kink-wf} \cite{markiewicz}.

We obtain waterfalls that compare fairly well with the experiments  
\cite{garcia,graf,chang}, although our perturbative scheme
underestimates their binding energy and broadening. In particular, our hot 
lines reproduce the loci of the BZ where the waterfalls are observed \cite{chang}
(shaded circles in Fig. \ref{fig:self-hs}). The waterfalls along the cuts A-C
(at binding energies $\approx 600$\,meV, $\approx 300$\,meV, and
$\approx 250$\,meV, respectively) correspond to the
nearly cross-shaped accumulation of the loci well inside the BZ in Fig. 
\ref{fig:self-hs}. In our scheme, these are due to C incipient order, which
also produces additional waterfalls along a square 
contour surrounding the $\Gamma$ point of the BZ. These are visible in panels B 
and C of Fig. \ref{fig4-kink-wf}, at approximately $700-800$\,meV. Their presence 
cannot be ascertained in Ref. \cite{chang}, where the data at higher binding energy 
are not reported. A reanalysis of the data is required to check for the
presence of these additional waterfalls.

On the other, hand both the C and S scattering are responsible for the 
dense occurrence of waterfalls near the $M$ points (cuts D-F in 
Fig. \ref{fig:self-hs}). However, as it is clear from panels D-F in Fig. 
\ref{fig4-kink-wf}, the waterfalls are shifted to lower binding energy in this 
region of the BZ and merge with the kinks. Morover, approaching the hot spots, 
the waterfall evolves into a rounding of the QP dispersions, 
with a spectral intensity vanishing as $\sqrt{\omega}$ \cite{abanov,chubukov,sulpizi}.
This rounding is reminiscent of the additional low-energy kinks observed in BSCCO 
\cite{multikink1,multikink2}, but not in LSCO, possibly due to a lower resolution.

\begin{figure}
\includegraphics[scale=0.3,angle=0]{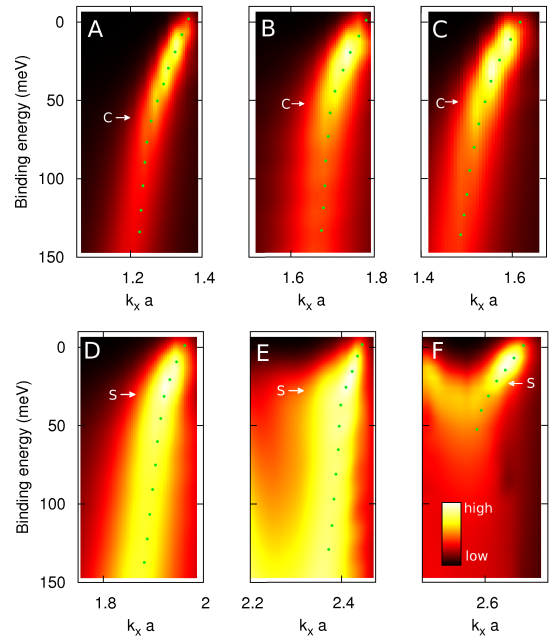}
\caption{(Color online) Same as in Fig. \ref{fig4-kink-wf}, but on a
narrower energy range. Although the structure of the kinks is 
due to both S and C CMs, the arrows mark the kinks due to (mostly) C or S scattering.}
\label{fig:spectra}
\end{figure}

\begin{figure}
\includegraphics[scale=0.28,angle=-90]{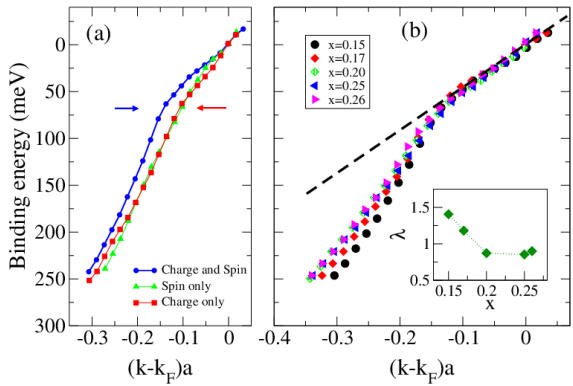}
\caption{(Color online) (a) QP dispersion along the $\Gamma X$ direction 
for LSCO at $x=0.15$ and $T=40$ K in the presence of both C and S CMs (solid circles, 
blue online) and for C (squares, red online) or S (triangles, green online) only. 
(b) Doping evolution of the kink. Both C and S CMs are considered, and $T=40$ K.
The dashed straight lines marks the low-energy dispersion. 
Inset: doping evolution of $\lambda\equiv(v_{HE}/v_F)-1$ (see text).}
\label{fig:sc_ch_sp}
\end{figure}

To follow the evolution of the kinks, we show in Fig. \ref{fig:spectra} the 
low-energy spectra, along the same cuts of Fig. \ref{fig4-kink-wf}.
In the nodal ($\Gamma X$) direction (cut A), the waterfall is well 
separated from the low-energy kink, which is more closely inspected in Fig. 
\ref{fig:sc_ch_sp} (a). Here we report the kinked dispersion separately due 
to C [(red) squares] and S [(green) diamonds] CMs, and to
the combination of both [(blue) circles].
The binding energy of the kink [(blu) arrow in Fig. \ref{fig:sc_ch_sp} (a)] is 
clearly set by the C CM [(red) arrow], at $\approx 60 - 70$\,meV, in 
good agreement with the experimental dispersion for LSCO at the same doping 
and temperature \cite{sahrakorpi}. The markedly propagating character of the
C CM (phonon-like away from the hot lines \cite{notaqbarra,sulpizi,becca})
makes its contribution to the kink rather sharp.

We emphasize that the characteristic energy of the C
CM is extracted from Raman experiments (Tab. \ref{tab}) and is not adjusted 
here by introducing additional modes at suitably chosen phonon frequencies.
On the other hand, the more diffusive S CM does not fix an energy scale and
rather renormalizes the bare QP dispersion over a broader energy range, affecting
the QP velocities far from the kink and making the kink more pronounced. Both 
C and S CMs must be simultaneously taken into account in order to quantitatively 
explain the kink.
The analysis of Raman spectra shows that the interaction mechanism switches from S 
to C CM with increasing $x$ \cite{noiraman}. This characterizes the doping evolution 
of the QP dispersion in the range $x=0.15-0.26$ \cite{sahrakorpi}. In Fig. 
\ref{fig:sc_ch_sp} (b), one sees that the S-vs-C switching produces no 
appreciable effect on the low-energy dispersion, which is 
determined cooperatively by the two CMs, so that the slope remains quite constant in 
the doping considered range [dashed line in Fig. \ref{fig:sc_ch_sp} (b)]. On the 
other hand, in the high-energy dispersion, where most of the effect of 
the C CM is exhausted and the variation of the slope is controlled by the the S CM, 
the dispersion becomes less steep with increasing doping, so that the high-energy 
QP velocity decreases. This is clearly observed looking at the inset in Fig. 
\ref{fig:sc_ch_sp} (b), where the parameter $\lambda\equiv(v_{HE}/v_F)-1$, measuring 
the strength of the (mostly S-mediated) interaction, is plotted vs. $x$. Although 
our analysis was limited to $x\ge 0.15$, 
extrapolating to the underdoped region the increasing strength of the 
S-mediated scattering (see Ref. \cite{noiraman}), we can also account for 
the observed \cite{sahrakorpi} increase of the parameter $\lambda$ (mainly due to
the increase of $v_{HE}$) below $x=0.15$.

Along different cuts in the BZ, the QP dispersion is differently affected by the 
two CMs, which may yield kinks (see Fig. \ref{fig:spectra}) 
at energies depending on the CM dispersion and on the position of the cut with 
respect to the hot lines. The kink due to the C CM moves to lower 
energies when the cut moves to the region where the C hot line approaches the 
Fermi surface (cuts B and C). On the other hand, the S hot lines track rather closely 
the Fermi surface,  giving rise to the above mentioned rounding of the dispersion at 
very low energy ($\lesssim 20$\,meV).
Along the cuts D-F, the C and S 
hot lines have intricate structures, which makes it difficult to distinguish the role 
of the two CMs in determining the mixed kink-waterfall structures. Nonetheless, 
by switching on and off the C and S couplings, we can state that the S CM plays the 
major role in this region of the BZ. 

The C-S cooperative behavior might be specific of LSCO, where the tendency to 
charge ordering seems to be more pronounced than, e.g., in YBCO, where the kinks
are more rounded. We also stress that our analysis only 
holds above $T_c$. Below $T_c$ the S CM changes and displays the peculiar resonance at 
$(\pi,\pi)$, which alters the shape of the kinks, producing a characteristic 
$s$-shaped dispersion in the antinodal regions \cite{norman1,norman2}. On the 
other hand, the rather broad and moderately coupled phonons should keep their 
effects (most pronounced around the nodal regions) even in the superconducting phase.

In conclusion, the salient aspects of ARPES experiments in LSCO are well reproduced 
by the {\it same} two (C and S) CMs previously obtained to fit 
Raman experiments. This solves for LSCO the long-standing issue whether the kinks 
are due to phonons or spin fluctuations: we reach the Solomonic conclusion that
both play a role. By the interplay of the two CMs, we can explain the highly 
non-trivial doping evolution of the low- and high-energy QP velocity along
the nodal ($\Gamma X$) direction, with $v_F$ 
almost doping independent and $v_{HE}$ decreasing with increasing doping 
(along with the suppression of the coupling with the S CM).
We predict the presence of multiple kinks (actually, a
kink and a low-energy rounding, analogous to 
those observed in BSCCO \cite{multikink1,multikink2}). We also
predict additional waterfalls at high binding energy, along a square contour 
around the $\Gamma$ point of the BZ.

Since, our analysis demonstrates the presence of {\it two} CMs, with characteristic 
wave vectors representative of stripe-like textures, our phenomenological model 
substantiates the presence of a competing C and S quasi-ordered phase  compatible with fluctuating stripes. 
The assessed relevant role of C and S CMs in LSCO also identifies them 
as candidate mediators of the pairing glue in these systems.

S.C., C. D.C., and M.G. acknowledge 
financial support from ``University Research Project'' of the ``Sapienza''
University n. C26A115HTN.


\begin{thebibliography}{99}

\bibitem{anderson}P. W. Anderson, Science {\bf 316}, 1705 (2007). 

\bibitem{scalapino} T. A. Maier, D. Poilblanc, and D. J. Scalapino, Phys. Rev. Lett. 
{\bf 100}, 237001 (2008).

\bibitem{hanke}W. Hanke, M. L. Kiesel, M. Aichhorn, S. Brehm, and E. Arrigoni, Eur. 
Phys. J. Special Topics {\bf 188}, 15 (2010).

\bibitem{abanov} Ar. Abanov, A. Chubukov, and J. Schmalian, Adv. Phys. {\bf 52}, 119 
(2003), and references therein.

\bibitem{varma}C. M. Varma, Phys. Rev. Lett. 75, 898 (1995); Phys. Rev. B {\bf 55}, 
14554 (1997), and references therein.

\bibitem{benfatto} L. Benfatto, S. Caprara, and C. Di Castro, Eur. Phys. J. B {\bf 17}, 
95 (2000). 

\bibitem{metzner} W. Metzner, D. Rohe, and S. Andergassen, Phys. Rev. Lett. {\bf 91},
066402 (2003).

\bibitem{kivelson} S. A. Kivelson, I. P. Bindloss, E. Fradkin, V. Oganesyan, J. M. 
Tranquada, A. Kapitulnik, 
and C. Howald, Rev. Mod. Phys. {\bf 75}, 1201 (2003) and references therein.

\bibitem{CDG} C. Castellani, C. Di Castro, and M. Grilli, Phys. Rev. Lett. {\bf 75}, 
4650 (1995).

\bibitem{garcia}For a review, see, e.g., 
D. R. Garcia and A. Lanzara, Adv. Cond. Mat. Phys. Volume 2010, Article ID 807412, 
doi:10.1155/2010/807412, and references therein.

\bibitem{bogdanov}
P. V. Bogdanov, A. Lanzara, S. A. Kellar, X. J. Zhou, E. D. Lu, W. J. 
Zheng, G. Gu, J.-I. Shimoyama, K. Kishio, H. Ikeda, R. Yoshizaki, Z. Hussain, and Z. X. 
Shen,  Phys. Rev. Lett. {\bf 85}, 2581 (2000).

\bibitem{zhou} X. J. Zhou, T. Cuk, T. Devereaux, N. Nagaosa, and Z.-X. Shen, 
``Angle-Resolved Photoemission Spectroscopy on Electronic Structure and 
Electron-Phonon Coupling in Cuprate Superconductors'', 
Handbook of High-Temperature Superconductivity: Theory and Experiment, edited by J. R. 
Schrieffer, (Springer, 2007), Pages 87-144. 

\bibitem{scalapino1} T. Dahm, V. Hinkov, S. V. Borisenko, A. A. Kordyuk, V. B. Zabolotnyy, J. Fink, B. BŸchner, D. J. Scalapino, W. Hanke and B. Keimer, Nat. Phys. {\bf 5}, 217 (2009).

\bibitem{norman1} A. V. Chubukov and M. R. Norman,  Phys.  Rev.  B {\bf 70}, 174505 
(2004).

\bibitem{graf}J. Graf, G.-H. Gweon, K. McElroy, S. Y. Zhou, C. Jozwiak, E. Rotenberg, A. 
Bill, T. Sasagawa, H. Eisaki, S. Uchida, H. Takagi, D.-H. Lee, and A. Lanzara, Phys. 
Rev. Lett. {\bf 98}, 067004 (2007).

\bibitem{chang} J. Chang, S. Pailh\'es, M. Shi, M. MŒnsson, T. Claesson, O. Tjernberg, J. 
Voigt, V. Perez, L. Patthey, N. Momono, M. Oda, M. Ido, A. Schnyder, C. Mudry, and J. 
Mesot, Phys. Rev. B {\bf 75}, 224508 (2007).

\bibitem{noiraman} S. Caprara, C. Di Castro, B. Muschler, W. Prestel, R. Hackl, M. 
Lambacher, A. Erb, S. Komiya, Y. Ando, and M. Grilli,
Phys. Rev. B {\bf 84 }, 054508 (2011).

\bibitem{notavandermarel}The presence of two collective modes has also been inferred from optical and experiments
in BSCCO systems in Refs. \cite{vandermarel,parmigiani}

\bibitem{vandermarel}E. van Heumen, E. Muhlethaler, A. B. Kuzmenko, H. Eisaki, W. Meevasana, M. Greven, and D. van der Marel, 
Phys. Rev. B {\bf 79}, 184512 (2009).

\bibitem{parmigiani} S. Dal Conte, C. Giannetti, G. Coslovich, F. Cilento, D. Bossini, T. Abebaw, F. Banfi, G. Ferrini, H. Eisaki, M. Greven, A. Damascelli, D. van der Marel, F. Parmigiani, Science {\bf 335}, 1600 (2012)

\bibitem{CDGZP} C. Castellani, C. Di Castro, and M. Grilli, Z. Phys. B {\bf 103}, 
137Ð144 (1997).

\bibitem{CDGJCPS}C. Castellani, C. Di Castro, and M. Grilli, J. Phys. Chem. Solids 
{\bf 59}, 1694 (1998).

\bibitem{loram}J. L. Tallon and J. W. Loram, Physica C {\bf 349}, 53 (2001).

\bibitem{perali}A. Perali, C. Castellani, C. Di Castro, and M. Grilli, Phys. Rev. B 
{\bf 54}, 16216 (1996).

\bibitem{andergassen}S. Andergassen, S. Caprara, C. Di Castro, and M. Grilli, Phys. Rev. 
Lett. {\bf 87}, 056401 (2001).

\bibitem{mmp} A. J. Millis, H. Monien, and D. Pines, Phys. Rev. B {\bf 42}, 167 (1990).

\bibitem{notaqbarra}Microscopic calculations \cite{becca} show that the C CM
essentially has a flattish phononic dispersion $\sim \omega_0$, which, near the C instability, is substantially softened in a
limited momentum region around $\gr{Q}_C$. The parameter  $\ovl{\gr{q}}_C$ sets the
width of this region, where a parabolic dispersion in the poles of Eq. (\ref{eq:CM_periodic}) extends from the energy scale $m_C$,
to $\sqrt{\ovl\Omega_C\Lambda_C}$. We assume that a similar momentum cutoff is present for the
S CM. 
One can also show \cite{tilman}
that $\ovl\Omega_C\approx \omega_0^2/t$
and $\Lambda_C\approx t$, yielding $\sqrt{\ovl\Omega_C\Lambda_C}\approx \omega_0$.

\bibitem{becca} F. Becca, M. Tarquini, M. Grilli, and C. Di Castro, Phys. Rev. B 
{\bf 54}, 12443 (1996).

\bibitem{tilman}S. Caprara, M. Grilli, C. Di Castro, and T. Enss, Phys. Rev. B {\bf  75}, 140505(R) (2007).

\bibitem{tranquada} J. M. Tranquada, B. J. Sternlieb, J. D. Axe, Y. Nakamura, and S.
Uchida, Nature (London) {\bf 375}, 561 (1995).

\bibitem{abbamonte}P. Abbamonte, A. Rusydi, S. Smadici, G. D. Gu, G. A. Sawatzky, and D. 
L. Feng, Nat. Phys. {\bf 1}, 155 (2005).

\bibitem{yamada} K. Yamada, C. H. Lee, K. Kurahashi, J. Wada, S. Wakimoto, S. Ueki, H. 
Kimura, Y. Endoh, S. Hosoya, G. Shirane, R. J. Birgeneau, M. Greven, M. A. Kastner, and 
Y. J. Kim, Phys. Rev. B {\bf 57}, 6165 (1998).

\bibitem{markiewicz} An explanation of the waterfalls observed in 
BSCCO in terms of an electronic (spin) generated self-energy 
was proposed by Susmita Basak, Tanmoy Das, Hsin Lin, J. Nieminen, M. Lindroos, R. 
S. Markiewicz, and A. Bansil, Phys. Rev. B {\bf 80}, 214520 (2009). In this 
case, however, only one (spin) mode was considered and the momenta where the 
waterfalls occur were not reported.


\bibitem{chubukov}  A. V. Chubukov, D. K. Morr, and A. Shakhnovich, Philos. Mag. B {\bf 74}, 563 (1996); 
A. V. Chubukov and D. K. Morr, Phys. Rep. {\bf 288}, 347 (1998).

\bibitem{sulpizi} S. Caprara, M. Sulpizi, A. Bianconi, C. Di Castro, and M. Grilli, Phys. 
Rev. B {\bf 59}, 14980 (1999).

\bibitem{multikink1} I. M. Vishik, W. S. Lee, F. Schmitt, B. Moritz, T. Sasagawa, S. Uchida, K. Fujita, S. Ishida, C. Zhang, T. P. Devereaux, and Z. X. Shen, Phys. Rev. Lett. {\bf 104}, 207002 (2010)

\bibitem{multikink2}S. Johnston, I. M. Vishik, W. S. Lee, F. Schmitt, S. Uchida, K. Fujita, S. Ishida, N. Nagaosa, Z. X. Shen, and T. P. Devereaux,
Phys. Rev. Lett. {\bf 108}, 166404 (2012).

\bibitem{sahrakorpi} S. Sahrakorpi, R. S. Markiewicz, Hsin Lin, M. Lindroos, X. J. Zhou, 
T. Yoshida, W. L. Yang, T. Kakeshita, H. Eisaki, S. Uchida, Seiki Komiya, Yoichi Ando, 
F. Zhou, Z. X. Zhao, T. Sasagawa, A. Fujimori, Z. Hussain, Z.-X. Shen, and A. Bansil, 
Phys. Rev. B {\bf  78}, 104513 (2008).


\bibitem{norman2} U. Chatterjee, D. K. Morr, M. R. Norman, M. Randeria, A. Kanigel, 
M. Shi, E. Rossi, A. Kaminski, 
H. M. Fretwell, S. Rosenkranz, K. Kadowaki, and J. C. Campuzano, Phys. Rev. B {\bf 75}, 
172504 (2007).


\end{thebibliography}

\end{document}